\documentclass[conference]{IEEEtran}

\usepackage[T1]{fontenc}
\usepackage{algorithm}
\usepackage{algorithmic}
\usepackage{blindtext, graphicx}
\usepackage{balance}
\usepackage{amsmath,lipsum}
\usepackage{nccmath}
\usepackage{cite}
\usepackage[short]{optidef}
\makeatletter
\newcommand*{\rom}[1]{\expandafter\@slowromancap\romannumeral #1@}

\makeatother

\usepackage[belowskip=-3.5pt,aboveskip=-1.5pt]{caption}
\let\subparagraph\paragraph
\usepackage[compact]{titlesec}

\titlespacing{\subsection}{1pt}{*1}{*1}

\begin{document}
\setlength{\abovedisplayskip}{0.2pt}
\setlength{\belowdisplayskip}{0.2pt}
\setlength{\abovedisplayshortskip}{0.2pt}
\setlength{\belowdisplayshortskip}{0.2pt}
\titlespacing*{\chapter}{0pt}{-50pt}{0pt}
\title{Q-Learning Based Aerial Base Station Placement for Fairness Enhancement in Mobile Networks}

\author{\IEEEauthorblockN{
Rozhina Ghanavi\IEEEauthorrefmark{1},
    Maryam Sabbaghian\IEEEauthorrefmark{1}, and
    Halim Yanikomeroglu\IEEEauthorrefmark{2}
  \IEEEauthorblockA{\IEEEauthorrefmark{1}
University of Tehran, Tehran, Iran, Email: \{rghanavi, msabbaghian\}@ut.ac.ir}
  \IEEEauthorblockA{\IEEEauthorrefmark{2}
  Carleton University, Ottawa, ON, Canada, Email: halim@sce.carleton}
  }
  \\[-4.0ex]}

\maketitle

\begin{abstract}
In this paper, we use an aerial base station (aerial-BS) to enhance fairness in a dynamic environment with user mobility. The problem of optimally placing the aerial-BS is a non-deterministic polynomial-time hard (NP-hard) problem. Moreover, the network topology is subject to continuous changes due to the user mobility. These issues intensify the quest to develop an adaptive and fast algorithm for 3D placement of the aerial-BS. To this end, we propose a method based on reinforcement learning to achieve these goals. Simulation results show that our method increases fairness among users in a reasonable computing time, while the solution is comparatively close to the optimal solution obtained by exhaustive search.\\

\end{abstract}

\IEEEpeerreviewmaketitle

\section{Introduction}
In recent literature, aerial base stations (aerial-BSs) have been proposed and studied as a way of tackling emergency situations or highly atypical load (traffic) conditions in wireless networks. A good example of the former is a natural disaster that could cause a terrestrial network to become nonoperational. To better cope with atypical traffic in space and time, the flexibility and agility of the network can be enhanced substantially by aerial-BSs; this would prevent over-engineering by eliminating the need for over-densification \cite{survey, survey1}. However, the performance of wireless networks with aerial-BSs is rather sensitive to the placement of the aerial-BSs. To mitigate this sensitivity, many studies have investigated the optimal placement of aerial-BSs in wireless networks \cite{jointal, test6, test8, test10, onnumber, backhaulaware, mount, optimal, flicons, wcncpaper, Deploy, dens, Traject}. 
It should be noted that, although the optimal placement of aerial-BSs is an important issue, the opportunities and challenges related to the use of aerial-BSs are not limited to their placement \cite{traj, fsobased, let}.

Ensuring high levels of fairness among users and a good cell-edge performance are some of the expectations of the next-generations of wireless networks. The proportional fairness model used in this paper results in a non-deterministic polynomial-time hard (NP-hard) problem. However, one of the most challenging aspects of solving NP-hard problems is finding a sufficiently accurate solution in a reasonable computing time. Moreover, the optimum aerial-BS locations obtained through computationally-expensive calculations for a snapshot of the dynamic network may become highly suboptimal as the topology changes, and accordingly the spatial load distribution evolves. For these reasons, reinforcement learning is an attractive candidate solution framework for the outlined aerial-BS positioning problem in a dynamic wireless network.\let\thefootnote\relax\footnote{This work is supported in part by Huawei Canada Research Centre.} We propose a solution based on reinforcement learning and compare it with other possible existing solutions in this area which are mainly heuristic algorithms. Not only do existing algorithms not promise the best results, they also yield solutions that are valid only for a snapshot of the system. We consider a wireless network composed of ground-BSs assisted by an aerial-BS to maintain a high level of fairness among users despite the user mobility as presented in Section \rom{2}.\textrm{B}. The backhaul link of the aerial-BS is an important constraint in the overall design; we assigned one of the ground-BSs to provide a backhaul link for the aerial-BS. Finally, we assume the presence of high-capacity fiber links to carry the data from ground-BSs to the core network.

The rest of this paper is organized as follows. Section \rom{2} presents the path loss and mobility models used in this paper. Section \rom{3} outlines the problem formulation and the proposed novel approach to solve it. Section \rom{4} presents the simulation results, and Section \rom{5} concludes the paper.

\section{System Model}
We consider the downlink of a wireless cellular network that includes some ground-BSs and several users. The network topology undergoes rapid changes due to user mobility. To ensure a high level of fairness in such dynamic environment, we exploit an aerial base station. This is beneficial to the network since the location of the aerial-BS can be adapted to the current status of the network. For the impartial assessment of the systems with and without the aerial-BS,  we keep the number of BSs constant in both scenarios. In fact, once the aerial-BS is added to the network, we use one of the ground-BSs as the backhaul of the aerial-BS. For the backhaul of the ground-BSs, we use fiber links which will not be congested. We denote the set of ground-BSs and the set of cellular users by $\mathcal{J}$ and $\mathcal{I}$, respectively. The cardinality of these sets are denoted by $J=|\mathcal{J}|$ and $I=|\mathcal{I}|$. We also show the association of the users to base stations by a binary parameter $U_{ij}$. If the $i$-th cellular user is associated to the $j$-th base station, $U_{ij}=1$, otherwise $U_{ij}=0$. The signal-to-interference-plus-noise ratio (SINR) for transmission from the $j$-th base station to the $i$-th user is denoted by $\gamma_{ij}$. The corresponding rate for this pair is presented by $R_{ij}$.

\subsection{Air-to-Ground Path Loss Model}
In the context of aerial-BS assisted terrestrial networks, path loss is widely considered the dominant term in the air-to-ground channel model. The path loss can be modeled as 
\begin{equation}
\textsf{PL }(\mathrm{dB})=\tau+\Lambda,
\end{equation}
where $\tau$, the free space path loss, is
\begin{equation}
\tau=20\log(d)+20\log(f)+20\log(\frac{4\pi}{c}),
\end{equation}
in which $f$ is the carrier frequency (in Hz), $c$ is the speed of light, and $d$ denotes the distance between the aerial-BS and user (in meters). The parameter $\Lambda$ is the average path loss which is obtained by taking the average of two cases of establishing a line-of-sight (LoS) or non-LoS (NLoS) link between the aerial-BS and the ground user. This can be expressed as
\begin{equation}
\Lambda=\textnormal{Pr}(\textnormal{LoS})\textsf{PL\textsubscript{LoS}}+\textnormal{Pr}(\textnormal{NLoS})\textsf{PL\textsubscript{NLoS}},
\end{equation}
where $\textnormal{Pr}(\textnormal{LoS})$ and $\textnormal{Pr}(\textnormal{NLoS})$ denote the probability of establishing a LoS or NLoS link between the aerial-BS and the ground user \cite{model1, model2}. These probabilities depend on the height of the aerial-BS and the elevation angle between the ground user and the aerial-BS. The losses due to LoS or NLoS links are denoted by $\textsf{PL\textsubscript{LoS}}$ and $\textsf{PL\textsubscript{NLoS}}$, respectively. The value of $\textsf{PL\textsubscript{LoS}}$ and $\textsf{PL\textsubscript{NLoS}}$ depend on the environment.

\subsection{User Mobility Model}
The prediction of users' traces based on real data has gained attention in various applications. In \cite{hiddenmarkov} an approach was presented to capture user mobility in cellular networks. This model, which is designed on the basis of real-life data, assigns specific destinations to each user as its point of interest. As users become more clustered in the network, ensuring fairness among them gets harder for ground-BSs. It is particularly situations like these that motivate the use of aerial-BSs in assisting the network. In this paper, we modify the model presented in \cite{hiddenmarkov} by considering $\nu$ social attractions and a random point as our destinations. In this model, we consider a plane that includes several users and $\nu$ places which are social attractions. We model the user movements by means of a Markov process which eventually clusters most of the users at these $\nu$ places. In fact, we assume that each user might select one of the $\nu$ places or a random point as its destination. These events are equiprobable with the probability of $\frac{1}{\nu+1}$. Once each user's destination is determined, we assume the user moves towards their destinations with a random speed between 0 and the maximum pedestrians speed, which is 1.3 m/s.

\subsection{Fairness Model}
Fairness plays an important role in the operation of a wireless network. It should be noted that fairness does not necessarily mean equal resource allocation. There are a number of fairness criteria which can be classified as quantitative or qualitative \cite{fairness in wireless}. The most common quantitative criteria are the Jain's index \cite{Jain}, entropy measure \cite{shanon}, unfairness measure \cite{test17} and Lorenz Curve \cite{test18}. The last two have received little attention in the fairness literature. The most common qualitative criteria are max-min fairness \cite{maxmin} and proportional fairness \cite{porpor}. The advantage of proportional fairness is that it ptovides a reasonable tradeoff between rate and fairness of the system. For this reason, in this paper we consider proportional fairness.


\section{Placing the Aerial-BS to Maximize Fairness}

\subsection{Optimization Problem}
Let us denote the rate obtained by the $i$-th user when it is connected to the $j$-th BS as
\begin{equation}
R_{ij}=b_{i}\log_2 (1+\gamma_{ij}),
\end{equation}
where $b_i$ is the assigned bandwidth to the $i$-th user. Without loss of generality, we assume $b_i=b$ for all users. The SINR can be expressed as $\gamma_{ij}=\dfrac{\tilde{P_{ij}}}{\sigma^2+\sum_{k=1}^{J}I_{ik}},$ where $\tilde{P_{ij}}$ is the received power from $j$-th BS at the $i$-th user and $\sigma^2$ is the zero-mean white Gaussian noise power. The term $I_{ik}$ illustrates the co-channel interference generated by the $k$-th BS on the $i$-th user. To maximize the proportional fairness, we have to solve the following optimization problem
\begin{alignat}{5}
&\!\max_{x_{\chi},y_{\chi},h_{\chi},U_{ij}}        &\qquad& \sum_{j=1}^{J}\sum_{i=1}^{I} {\log(R_{ij}U_{ij})},\\ 
&\text{s.t.} &      & \sum_{i=1}^{N'}{R_{i\chi}} \leq c_{\zeta},\\
&                  &      & R_{ij}U_{ij}\geq R_{\textnormal{min}}U_{ij}, \quad{\forall}{i \in \mathcal{I}} ,{\forall}{j \in \mathcal{J}},\\
&                  &      & \sum_{i=1}^{N'} {P_{i\chi}}\leq P_{\textnormal{max}},\\
&                  &      & \sum_{j=1}^{J} {U_{ij}}=1,\\
&                  &      & U_{ij} \in \{0,1\}, \qquad {\forall}{i \in \mathcal{I}} ,{\forall}{j \in \mathcal{J}},
\end{alignat}
where $\chi$ denotes the aerial-BS and $N'$ is the number of users assigned to the aerial-BS. The coordinates $x_{\chi},y_{\chi},h_{\chi}$ denote the 3D location of the aerial-BS. These parameters can vary in the following ranges $[x_\textrm{min},x_\textrm{max}]$, $[y_\textrm{min},y_\textrm{max}]$, and $[h_\textrm{min},h_\textrm{max}]$. $U_{ij}$ is determined by maximum SINR criteria. Constraint (6) guarantees that the sum-rate of the aerial-BS does not exceed $c_{\zeta}$ which is the maximum capacity of its backhaul. Constraint (7) illustrates the minimum rate requirements for the users. Constraint (8) presents the power limit for the aerial-BS. Constraint (9) assigns each user to only one BS. The last constraint shows that the user association coefficient can only take binary values.

\subsection{Efficient Placement of the Aerial-BS}The optimization problem of (5) can be reduced to an NP-hard problem. Since the problem is NP-hard, deriving a closed form solution is not feasible. This fact, combined with the dynamic nature of the network, motivate us to exploit alternative solutions. The solution has to be adapted quickly to dynamic network alterations. Taking these points into account, approaches based on reinforcement learning provide an appropriate platform to solve this problem. We consider simulated annealing (SA) Q-learning whose convergence speed is reasonable \cite{saqlearning}. The SA algorithm is an optimization algorithm which modifies the solution based on the Metropolis criterion. In SA-Q-learning, Q-learning is used to obtain the optimal procedure. Then, the Metropolis criterion, which is the core of the SA algorithm, is applied to select between the policy $\pi$, which is the exploration and exploitation of the action in the learning procedure. In \cite{saqlearning}, it has been shown that SA-Q-learning outperforms $\epsilon$-greedy Q-learning in terms of convergence speed.
\setlength{\textfloatsep}{0pt}
\begin{algorithm}

\caption{Proposed Algorithm}
 \begin{algorithmic}[1]

  \STATE Initialize $Q$ matrix as the previous session matrix in which the aerial-BS was being used;
  \WHILE {System is running}
  \STATE Users are moving in the area for $t\textsubscript{min}$;
  \STATE Repeat for each episode;
  \STATE Repeat for each step in the episode;
  \STATE Select a random action $a_{r}$ in $A(s)$;
  \STATE Select an action $a_{p}$ in $A(s)$ which maximizes reward;
  \STATE Generate a random value $\epsilon \in (0,1)$;
  \IF {($\epsilon < \exp{\frac{Q(s,a_{r})-Q(s,a_{p})}{\psi}}$)}
  \STATE Take the action $a_{r}$;
  \ELSE 
  \STATE Take the action $a_{p}$;
  \ENDIF
  \STATE
   $Q(s_{t},a_{t})=\alpha[r_{t+1}+\eta\textnormal{max}\{Q(s_{t+1},a_{t+1})\}-Q(s_{t},a_{t})]$
  \STATE Update the state of the aerial-BS;
  \STATE Update the assignment of users to BSs $(U_{ij})$ by maximum SINR constraint;
  \STATE Update $\psi$ as follows, $\psi_{t+1}=\lambda \psi_{t}$.
  \ENDWHILE
 \end{algorithmic}
 \end{algorithm}
 
 The Q-learning algorithm is a model free reinforcement learning algorithm. It is notable that Q-learning is an off-policy algorithm, meaning that it learns to optimize the target function while following the action policy. In these methods no matter what sequence of actions the agent takes, it will converge to the optimum point if it has enough learning time. This algorithm is a Markov decision process that consists of the following elements: states, policy, actions, transition probabilities, reward, and knowledge metric. The set of states, $S=s\textsubscript{1},s\textsubscript{2},...,s_{v},$ describes the system. The policy $\pi$ determines the action to be taken in the current state of the system. We show the set of actions by $A=a\textsubscript{1},a\textsubscript{2},...,a_{w}.$ The transition from one state to another occurs according to a specific action. The probability of this event is called transition probability. Each transition entails a specific reward. The performance of the policy is measured by the knowledge matrix, $Q$. Each element of the Q-matrix is associated with one of the state-action pairs. In the learning phase, the matrix is first initialized with proper values. Once we run the algorithm, the matrix components are updated, as several states are visited. In each state, an action is exploited or explored which leads to a transition to the next state. The reward involved in each transition is used to update the Q-matrix. At the end of the learning phase, the action whose Q-factor has the highest value for each state is selected as the best action. In our problem, the geographical coordinates of the aerial-BS determine the state of the system. In fact, we discretize the 3D flying zone of the aerial-BS into smaller cubes whose side length is $\upsilon$. The center of each cube is the state of the system. The action is the movement of the aerial-BS $\upsilon$ meters towards any of the six faces of the cube. Defining a proper reward in the learning algorithm is essential to solving the problem of (5). We consider a reward as

\begin{equation}
    r_{t}=r_{t}^+-r_{t}^-
\end{equation}
where $t$ denotes the state number. The terms of $r_t$ are
\begin{alignat}{2}
& r_{t}^+=\Theta_{t}-\Theta_{t-1}+\omega_{t}-\omega_{t-1},\\
& r_{t}^-=\delta_{1}(\beta_{t}-\beta_{t-1}),
\end{alignat}
where $\Theta_t$, $\beta_t$, and $\omega_t$ are defined as follows
\begin{alignat}{3}
&\Theta_{t}=  \sum_{j=1}^{J}\sum_{i=1}^{I} {\log(R_{ij}U_{ij})},\\
      & \beta_{t}=\eta_{1}\mathcal{H}[R_{i\chi}-c_{\zeta}],\\
    & \omega_{t} = \sum_{j=1}^{J}\sum_{i=1}^{I} {U_{ij}\gamma_{ij} \mathcal{H}[R_{ij}-R\textsubscript{min}]}.
\end{alignat}

In fact, $\Theta_t$ is the fairness achieved at the $t$-th iteration, and $\omega_t$ is the sum of SINRs for the users who satisfy the minimum rate requirement at the $t$-th iteration. The term $\beta_t$ presents the penalty if the backhaul capacity constraint of the aerial-BS is violated, and $\mathcal{H}[.]$ is the step function. Thus, the reward presented in (11) reflects the objective function and constraints (6) and (7).

The $Q$ matrix at the $t$-th time interval is modified as
$Q(s_{t},a_{t})=\alpha[r_{t+1}+\eta \textnormal{max}\{Q(s_t{t+1},a_{t+1})\}-Q(s_{t},a_{t})],$ where $\alpha$ is the decreasing learning rate and $\eta$ is the discount factor.
In SA-Q-learning, the state transition algorithm is
\begin{equation}
    f(\lambda1 \rightarrow \lambda2) = \begin{cases}
               a_{r},               & \mbox{if} \qquad \epsilon < \exp{\frac{Q(s,a_{r})-Q(s,a_{p})}{\psi}}\\
               a_{p}, & \text{otherwise,}
           \end{cases}
\end{equation}
where $\lambda1$ and $\lambda2$ represent the current and future states, respectively. The parameter $\psi$ is a decreasing parameter in the process, and $\epsilon \in (0,1)$ is a random number. Random and optimum actions for each state are presented by $a_r$ and $a_p$. The learning algorithm requires specific conditions to converge to the global optimum point, which can be found in \cite{dayan}. These conditions are all satisfied is our scenario. Learning parameters are presented in Table \rom{1}. These values are tuned through a simulation-based search. Algorithm 1 shows the SA-Q-learning algorithm that we used.
 
\begin{figure}
  \includegraphics[width=0.49\textwidth]{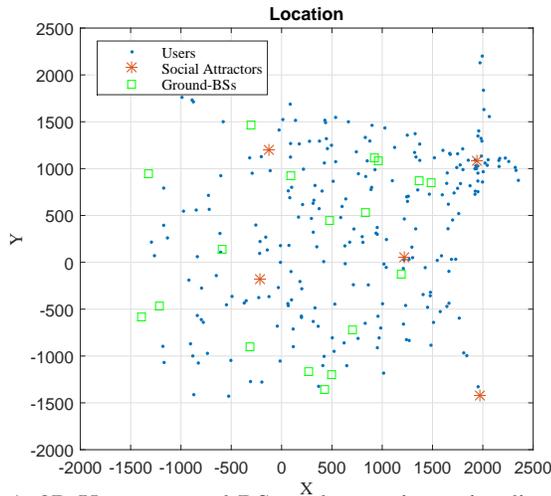}
  \centering
  \caption{2D Users, ground-BSs and attracting point distribution after moving with the proposed mobility model at $t=25$ min.}
\end{figure}

\begin{figure}
  \includegraphics[width=0.5\textwidth]{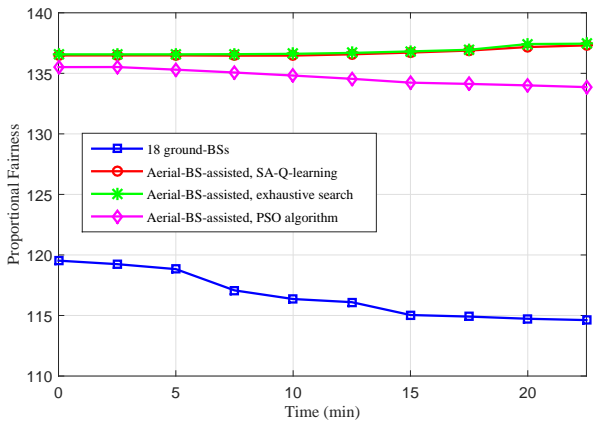}
  \centering
  \setlength\belowcaptionskip{+10pt}
  \caption{Proportional fairness from (5) for traditional ground-BS system and proposed system assisted with an aerial-BS.}
\end{figure}

\section{Simulation Results}
We consider the problem of an urban cellular network, where 18 ground-BSs are positioned according to a binomial point process (BPP) in a 4 km x 4 km region. A random number of users in the interval $[m\textsubscript{min}, m\textsubscript{max}]$ (with uniform distribution) are also placed in the same region using another BPP; the two BPPs are independent. At t = 0, the users start moving according to the mobility model presented in Section \rom{2}.\textrm{B}. In this network, $\nu$ social attracting points are placed on the basis of another independent BPP. As time passes, some users tend to cluster around the attracting points, while others move randomly. Users associations to BSs change after each iteration. Simulation parameters are shown in Table \rom{1}. Fig. 1 presents the user and ground-BS distribution at $t=25$ mins. As we can see, some users tend to cluster around the attracting points, denoted by stars in Fig. 1, while others move randomly.

\begin{figure}
  \includegraphics[width=0.5\textwidth]{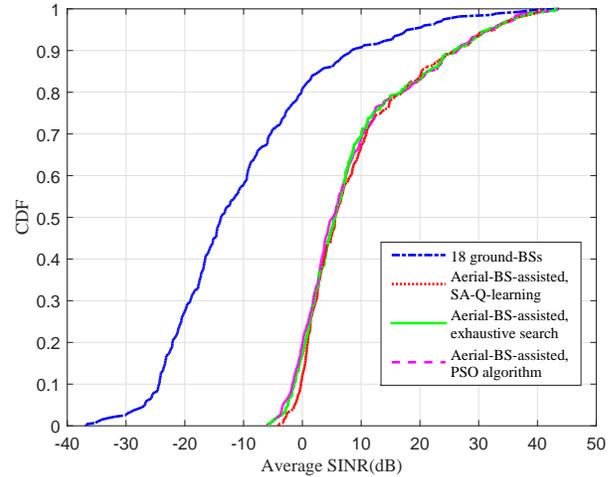}
  \centering
  \caption{CDF of average SINR of users for the traditional ground-BS system and proposed system assisted with an aerial-BS.}
\end{figure}
We compare the performance of a traditional system with 18 ground-BSs with an aerial-BS assisted system with 17 ground-BSs and one aerial-BS. For the optimal placement of the aerial-BS, we considered 3 methods for solving the mentioned problem, an exhaustive search method, our proposed SA-Q-learning method, and a particle swarm optimization (PSO) method \cite{onnumber}. Fig. 2 presents the proportional fairness, and Fig. 3 shows the cumulative distribution function (CDF) of average SINR for these scenarios. As presented, our method obtains the proportional fairness near to the results from exhaustive search which is higher than that of PSO and significantly higher than that of the traditional system. It is notable that the proportional fairness from (5) is a logarithmic function of rate; hence from Fig. 2, the rate improvement is impressive. In Fig. 2 the proportional fairness from traditional ground-BSs system is decreasing over time while users are moving in accordance with the the mobility model presented in Section \rom{2}.\textrm{B} and cluster. This shows the effectiveness of our proposed model since it can attain fairness in any situation. As we can see in Fig. 3, SA-Q-learning and PSO perform similar to exhaustive search. The results from Fig. 3 indicate that cell-edge performance from the proposed method has improved 30 dB. Other simulation results show convergence rate of both SA-Q-learning and $\epsilon$-greedy Q-learning are good for our use case. They also show our SA-Q-learning method can reach to the optimum point in an acceptable computing time. The simulation results conclude that the main benefit of our solution is its fast adaptation to continuous changes of the network topology while achieving the optimum solution.

\begin{table}[t!]
  \renewcommand{\arraystretch}{1.3}
  \centering
  \caption{Simulation Parameters}\label{table1}
    \begin{tabular}[t]{|c |c |c |c |} 
      \hline
      \scriptsize\textbf{Parameter} & \scriptsize\textbf{Value} & \textbf{Parameter} & \scriptsize\textbf{Value} \\  
      \hline
      $(P\textsubscript{LoS}, P\textsubscript{NLoS})$ & (1 dB, 20 dB)&
      $f$ & 2 GHz  \\
      \hline
      
      $(m\textsubscript{\textnormal{min}}, m\textsubscript{\textnormal{max}})$ & (200, 300) &
      BW & 20 MHz \\
      \hline
      
      $(h\textsubscript{\textnormal{min}}, h\textsubscript{\textnormal{max}})$ & (25 m, 525 m) &
      $R_{\textnormal{min}}$ & 0 \\
      \hline
      
      $(x\textsubscript{\textnormal{min}}, x\textsubscript{\textnormal{max}})$ & (-2000 m, 2000 m) &
      $P_{\textnormal{max}}$ & 49 dBm \\
      \hline
      
      $ (y\textsubscript{\textnormal{min}}, y\textsubscript{\textnormal{max}})$ & (-2000 m, 2000 m) &
      $\nu$ & 5 \\
      \hline
      $\psi$ & 10 &
$\lambda$ & 0.99  \\
\hline
$(\eta_{1}, \delta_{1})$ & (1000, 100) &
$t\textsubscript{min}$ & 2.5 mins\\
\hline
$\eta$ & 0.9 &
$\upsilon$ & 10 m\\
\hline 
    \end{tabular}
    \end{table}
\section{Conclusion}
In this paper, we proposed a method to achieve fairness among users in an aerial-BS empowered terrestrial network with user mobility. To obtain the optimal placement of the aerial-BS when the network is dynamic, we used SA-Q-learning method. We showed that the aerial-BS can significantly improve the proportional fairness among users. We also showed that for the aerial-BS assisted network, the SA-Q-learning method is the optimal performance achieved by exhaustive search while at the same time being faster and adaptable to new situations in the system.



\section*{Acknowledgement}
The authors would like to thank Elham Kalantari and Abbas Yongacoglu from the University of Ottawa for their insights and support in making this work possible.

\balance

\end{document}